\documentclass[journal]{IEEEtran}
\usepackage{ifpdf}

\usepackage{cite}

\ifpdf
   \usepackage[pdftex]{graphicx}
   \graphicspath{{figures/}}
  \DeclareGraphicsExtensions{.pdf,.jpeg,.png,.eps,.jpg}
\else
\fi

\usepackage{amsmath}
\usepackage{amssymb}

\newcommand{\vc}[1]{{\bf #1}}

\newcommand{\ma}[1]{{\bf #1}}

\newcommand{\myexp}[1]{e^{{#1}}}

\DeclareMathOperator*{\argmax}{argmax} 

\newtheorem{proposition}{Proposition}

\newtheorem{property}{Property}

\usepackage{todonotes}

\begin{document}
%
\title{Augmented Slepians: Bandlimited Functions that Counterbalance Energy in Selected Intervals}
%
%
%

\author{Robin~Demesmaeker,~
        Maria~Giulia~Preti,~\IEEEmembership{Member,~IEEE,}
        and~Dimitri~Van~De~Ville,~\IEEEmembership{Senior Member,~IEEE}
\thanks{R.\ Demesmaeker, M.\ G.\ Preti, D.\ Van De Ville are with the Institute of Bioengineering/Center for Neuroprosthetics, Ecole Polytechnique F\'ed\'erale de Lausanne, and the Department of Radiology and Medical Informatics, University of Geneva, Switzerland.}
\thanks{Manuscript received xxx; revised xxx; accepted xxx; published xxx.}}

%
%

\markboth{Submitted to IEEE Transactions on Signal Processing}%
{Demesmaeker \MakeLowercase{\textit{et al.}}: Generalized Slepian Functions }
%



\maketitle


\begin{abstract}
Slepian functions provide a solution to the optimization problem of joint time-frequency localization. Here, this concept is extended by using a generalized optimization criterion that favors energy concentration in one interval while penalizing energy in another interval, leading to the ``augmented'' Slepian functions. Mathematical foundations together with examples are presented in order to illustrate the most interesting properties that these generalized Slepian functions show. Also the relevance of this novel energy-concentration criterion is discussed along with some of its applications. 
\end{abstract}


%
\IEEEpeerreviewmaketitle

\section{Introduction}
%
%
%
%
\IEEEPARstart{H}{eisenberg}'s uncertainty principle states that the energy of a signal can never be strictly localized both in the temporal and the Fourier domain. In a series of seminal papers, Slepian, Pollak, and Landau~\cite{Slepian.1961,Landau.1961,Landau.1962,Slepian.1964} study the case where maximal energy concentration on a selected interval is sought for a band-limited function. They show that the solution can be found from an integral eigenvalue equation where eigenvalues indicate energy concentration in the selected interval, and eigenfunctions define a basis that is orthonormal on $\mathbb{R}$, and orthogonal on the selected interval. The sum of the eigenvalues---which exhibit a striking phase transition between high and low energy concentrations---corresponds to the time-bandwidth product (a.k.a. the Shannon number) and characterizes the dimensionality of the linear space of band-limited functions associated to an interval with given width. The functions defined by Slepian \emph{et al.}, called Slepian functions onward, are known as prolate spheroidal wave functions and have a number of elegant properties and applications, including band-limited extrapolation. They have also been extended for other domains such as their spherical counterparts with applications in geophysics~\cite{Simons_2006}.

It is straightforward to extend the Slepian construction for two or more intervals in the temporal domain. In such case, the solution maximizes the energy simultaneously in all intervals. However, in some applications, it can be useful to be able to specify intervals that are counteracting; i.e., when one wants to obtain functions that are maximally concentrated in one interval, while being minimally concentrated in another one. Therefore, in this paper, we generalize Slepian functions by pursuing band-limited functions that not only maximize energy concentration in one interval, but are also penalized by their energy concentration in another one. We demonstrate that the solution can still be found from an integral eigenvalue equation where the eigenvalues indicate the \emph{difference} in energy concentration between both intervals. The eigenspectrum reveals two phase transitions with corresponding time-bandwidth products. The eigenfunctions are approximately orthogonal on the selected intervals. The interaction between both intervals makes the solution effectively different from combining solutions for the intervals separately. 

The paper is organized as follows. After a short review of 1-D Slepian theory in Sec.~\ref{sec:slepfun}, we introduce the proposed generalization (Sec.~\ref{sec:penslepfun}). We provide the mathematical foundations together with instructive 1D examples and several properties of the ``augmented'' Slepian functions. To conclude, we discuss possible applications of this novel view on energy localization in domains such as signal recovery and data analysis.

\section{Slepian Functions\label{sec:slepfun}}
Slepian and colleagues were the first to propose an elegant solution to the problem of finding continuous-domain functions that are band-limited, but  with maximal energy concentration in an interval. We briefly review the Slepian theory, highlighting those aspects that are important for the generalization. 

Let us start by introducing the Hilbert space of square-integrable functions $L_2(\mathbb{R})$ with associated inner product
\begin{equation}
  \left< f, g \right>_\mathbb{R} = \int_{\mathbb{R}} f(t) \bar{g}(t) dt,
\end{equation}
where $\bar{ \cdot}$ is the complex conjugate. The Fourier transform (and its inverse) is defined as
\begin{eqnarray}
 F(\omega) & = & \int_\mathbb{R} f(t)\exp(-i\omega t)dt, \\
  f(t) &=& \frac{1}{2\pi}\int_\mathbb{R} F(\omega)\exp(i\omega t)d\omega.
\end{eqnarray}
The Slepian design problem can be formulated as finding the band-limited function $g(t)$ that maximizes the energy concentration in an interval. The temporal interval is chosen $[-T,+T]$ and thus centered around the origin with a half width of $T$. The spectral band-limit is specified as $[-W,+W]$ where $W$ indicates the one-sided bandwidth. The following optimization problem for maximizing the energy concentration can then be written: 
\begin{equation}
  \lambda = \max_{g(t)\in {\mathcal B}_W} \frac{ \int_{-T}^{+T} \left|g(t)\right|^2 dt}{ \int_\mathbb{R} \left|g(t)\right|^2 dt}, 
\end{equation}
where $\mathcal{B}_W$ is the space of band-limited functions in $[-W,+W]$. As shown by Slepian \emph{et al.}, this criterion can be reformulated in the Fourier domain as
\begin{equation*}
\frac{\frac{1}{2\pi}\int_{-W}^{+W}\int_{-W}^{+W} G(\omega)\bar{G}(\omega') \overbrace{\frac{1}{2\pi} \int_{-T}^{+T} \myexp{-i(\omega' - \omega) t} dt}^{=D(\omega'-\omega)} d\omega d\omega'}{\frac{1}{2\pi}\int_{-W}^{+W} G(\omega) \bar{G}(\omega) d\omega},
\end{equation*}
where the kernel
\begin{equation}
  D(\omega) = \frac{1}{2\pi} \int_{-T}^{+T} \myexp{-i\omega t} dt = \frac{\sin(T \omega)}{\pi\omega}
\end{equation}
is the scaled Fourier transform of the indicator function of the interval $[-T,+T]$. Therefore, maximizing the energy concentration leads to an equivalent integral eigenvalue equation in the Fourier domain: 
\begin{equation}
 \int_{-W}^{+W} {\frac{ \sin T(\omega-\omega^\prime)}{\pi(\omega-\omega^\prime)}} G(\omega')d\omega' =  \lambda G(\omega)\label{eq:freq_dom}. 
\end{equation}
This equation can be written in its canonical form by replacing $G(\omega)=\psi(\omega/W)$ and the change of variables $\omega=W\xi$: 
\begin{equation}
 \int_{-1}^{+1} {\frac{ \sin TW(\xi-\xi')}{\pi(\xi-\xi')}}  \psi(\xi')d\xi' =  \lambda \psi(\xi).
\label{eq:freq_dom_norm}
\end{equation}
 
Since the kernel in this homogeneous Fredholm equation of the second kind is symmetric positive definite, the integral operator is compact and its solutions $\lambda_k$, $\psi_k$, $k\in \mathbb{N}$, are countable where the eigenvalues $\lambda_k$ are positive (and tend to zero), and the real-valued eigenfunctions $\psi_k$, known as prolate spheroidal wave functions (PSWF), hereafter also referred to as Slepian functions, form an orthogonal basis of $L_2([-1,1])$. 
The PSWF can be extended to build an orthogonal basis of $L_2(\mathbb{R})$ by defining Eq.~(\ref{eq:freq_dom_norm}) for all $\xi\in \mathbb{R}$. This leads to the double orthogonality property 
\begin{eqnarray}
  \left< \psi_k, \psi_l \right>_{\mathbb R} & = & \delta_{k-l}, \\
  \left< \psi_k, \psi_l \right>_{[-1,+1]} & = & \lambda_k \delta_{k-l},
\end{eqnarray}
where $\delta_k$ is the Kronecker delta.\\
In addition, any PSWF also satisfies the following intriguing Fourier property:
\begin{equation}
   \psi_k(t) = \frac{1}{\mu_k} \int_{-1}^{+1} \psi_k(\xi) \myexp{i TW t\xi} d\xi,
   \label{eq:Fourierid}
\end{equation}
where $\mu_k\in\mathbb{C}$ is a scaling factor up to which the PSWF has the same shape as its Fourier transform in the interval $[-1,+1]$. This property plays a key role in relating the PSWF to the prolate differential equation that justifies their name and provides an alternative numerical procedure for their computation. 

An important feature of the Slepian basis is the Shannon number $N_\text{Shannon}$, which is given by the sum of all eigenvalues. It can easily be shown that this number only depends on the time-bandwidth product $2TW$: 
\begin{equation}
N_{\text{Shannon}}=\sum_{k=0}^\infty\lambda_k=\int_{-W}^{+W} \lim_{\xi'\to\xi}D(\xi-\xi')d\xi=\frac{2TW}{\pi} \label{eq:shannon}
\end{equation}
Since the characteristic spectrum shows a step-like behaviour with eigenvalues either close to $1$ or $0$ separated by a narrow transition band, $N_\text{Shannon}$ moreover approximately represents the number of eigenfunctions that are well concentrated within the selected region of interest. Therefore, it is also a measure for the dimension of the subspace spanned by the band-limited  functions that are well localized.

The Slepian construction can easily be extended for an interval that is not centered at the origin. In such case, the following modified properties hold: 
\begin{proposition}[Translated temporal interval]\label{prop:translated_temporal_interval}
The Slepian design for a translated temporal interval $[-T+P,+T+P]$ satisfies the following Fourier property
\begin{equation}
   \psi_k(t) = \frac{\myexp{iPWt}}{\mu_k} \int_{-1}^{+1} \psi_k(\xi) \myexp{i (PW+TWt)\xi} d\xi,
\end{equation}
and corresponding integral eigenvalue equation:
\begin{equation}
   \psi_k(\xi) = \underbrace{\frac{2\pi}{\left|\mu_k\right|^2 TW}}_{=1/\lambda_k} \int_{-1}^{+1} \psi_k(\xi') {\myexp{iPW(\xi-\xi')} \frac{\sin(TW(\xi-\xi'))}{\pi(\xi-\xi')}} d\xi'. 
\end{equation}
\end{proposition}
Proof: See Appendix \ref{app:translated_temporal_interval}. 

One can ask whether the Slepian design can be further extended to a union of intervals. The answer is affirmative from a point-of-view of the construction of the energy-concentration criterion to be optimized. Specifically, let us consider the union of intervals 
$$
   \mathcal{S} = \bigcup_{n=1}^N [-T_n+P_n,+T_n+P_n],
$$ 
for which criterion to be maximized is
\begin{equation}
  \lambda = \max_{g(t)\in {\mathcal B}_W} \frac{ \int_\mathcal{S} \left|g(t)\right|^2 dt}{ \int_\mathbb{R} \left|g(t)\right|^2 dt},
\end{equation}
which can be turned into the equivalent integral eigenvalue equation
\begin{equation*}
  G_k(\omega) = \frac{1}{\lambda_k} \int_{-W}^{+W} \underbrace{\sum_{n=1}^N\myexp{iP_n(\omega-\omega')} \frac{\sin(T_n(\omega-\omega'))}{\pi(\omega-\omega')}}_{=D(\omega-\omega')} G_k(\omega') d\omega'.
\end{equation*}
Unfortunately, the Fourier property (\ref{eq:Fourierid}) no longer holds. Therefore, we need to explicitly define the temporal domain version of these functions as
\begin{equation}
  g_k(t) = \frac{1}{2\pi} \int_{-W}^{+W} G_k(\omega) \myexp{i t\omega} d\omega,
\end{equation}
to which we will refer to as the ``Slepian functions'' since they do not necessarily correspond to PSWFs. 
Using the normalization $\left\|g_k\right\|^2=1$ and thus $\left\|G_k\right\|^2=2\pi$ (Parseval identity), we can still easily prove the double orthogonality property in the temporal domain.
\begin{proposition}[Orthogonality of Slepian functions for union of intervals]
\label{prop:ortho}
The Slepian functions $g_k$, $k\in \mathbb{N}$, associated to the union of intervals $\mathcal{S}$, satisfy the following double orthogonality property:
\begin{eqnarray}
  \left< g_k, g_l \right>_{\mathbb R} & = & \delta_{k-l}, \\
  \left< g_k, g_l \right>_{\mathcal{S}} & = & \lambda_k \delta_{k-l}.
\end{eqnarray}
\end{proposition}
The proof is given in Appendix~\ref{app:ortho}.

\section{Augmented Slepian Functions\label{sec:penslepfun}}

\subsection{Design}
The PSWF introduced in the previous section are driven by maximizing energy concentration in a chosen interval. This implies that the energy is minimized everywhere else since 
\begin{equation*}
 \max_{g(t)\in {\mathcal B}_W}\left(\frac{ \int_\mathcal{S} \left|g(t)\right|^2 dt}{ \int_\mathbb{R} \left|g(t)\right|^2 dt}\right) 
 = \max_{g(t)\in {\mathcal B}_W}\left(1 - \frac{ \int_{\mathcal{S}^c} \left|g(t)\right|^2 dt}{ \int_\mathbb{R} \left|g(t)\right|^2 dt}\right),
\end{equation*}
where $\mathcal{S}^c=\mathbb{R}\backslash \mathcal{S}$ is the complement of $\mathcal{S}$.

\begin{figure}[t]
\centering
\includegraphics[width = .9\linewidth]{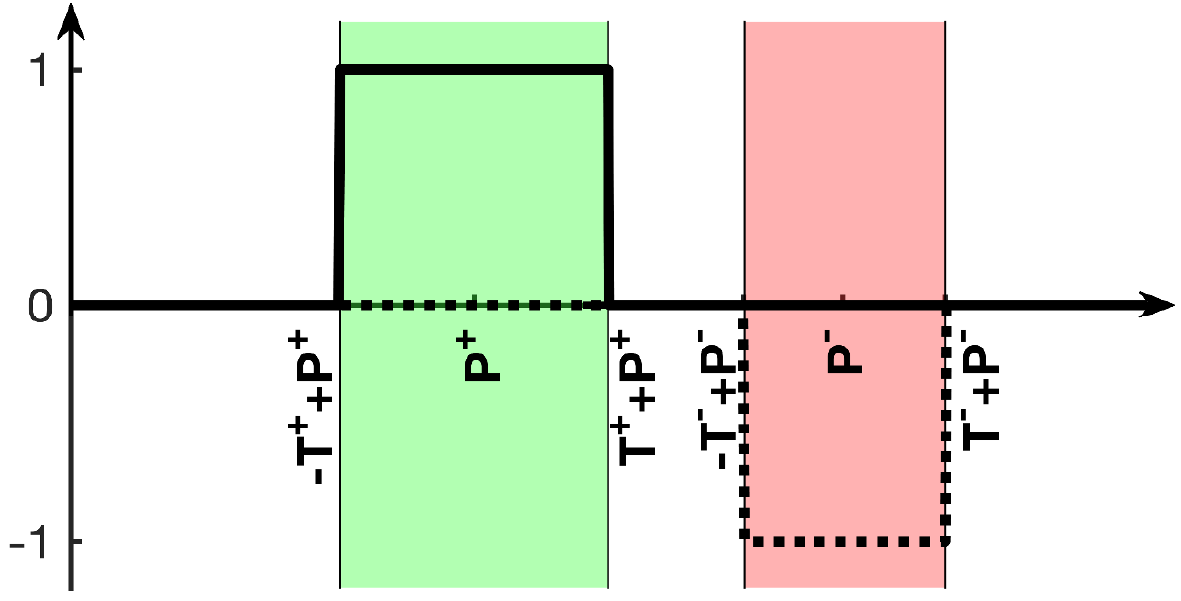}
\caption{Schematic presentation of the concept of selecting two types of intervals. The one in green (``positive'', by the plain indicator function) and the one in red (``negative'', by the dashed indicator function). The criterion that is maximized will be the difference between the energies in the green and the red interval, respectively, normalized with respect to the energy on the real line. 
\label{fig:setup_intervals}}
\end{figure}

Here, we propose to introduce explicitly the notion of a second type of interval. As illustrated in Fig.~\ref{fig:setup_intervals}, we want to find the band-limited functions that maximize the energy concentration in the green interval while minimizing it in the red interval, where both can be chosen by the user. Consequently, the energy in the two intervals will be counterbalanced. This additional freedom in the design leads to what we term as ``augmented Slepians''. Mathematically, the criterion to be maximized is defined as
\begin{equation}
  \lambda = \max_{g(t)\in {\mathcal B}_W}\left(\frac{ \int_\mathcal{S^+} \left|g(t)\right|^2 dt -  \int_\mathcal{S^-} \left|g(t)\right|^2 dt}{ \int_\mathbb{R} \left|g(t)\right|^2 dt}\right),
  \label{eq:criterion_augmented}
\end{equation}
where $\mathcal{S}^+$ and $\mathcal{S}^-$ are two (disjoint) unions of intervals: 
\begin{eqnarray*}
   \mathcal{S}^+ & = & \bigcup_{n=1}^{N^+} [-T^+_n+P^+_n,+T^+_n+P^+_n], \\
   \mathcal{S}^- & = & \bigcup_{n=1}^{N^-} [-T^-_n+P^-_n,+T^-_n+P^-_n].
\end{eqnarray*}
By turning the criterion (\ref{eq:criterion_augmented}) in the Fourier domain, we find the corresponding integral eigenvalue equation for a generalized kernel: 
\begin{eqnarray}
 \int_{-W}^W D(\omega-\omega^\prime)G(\omega^\prime)d\omega^\prime&=& \lambda G(\omega),\label{eq:integral_augmented}
\end{eqnarray}
where
\begin{eqnarray*}
D(\omega) & = & \sum_{n=1}^{N^+}\myexp{iP^+_n\omega} \frac{\sin(T^+_n\omega)}{\pi\omega} \\
& & - \sum_{n=1}^{N^-}\myexp{iP^-_n\omega} \frac{\sin(T^-_n\omega)}{\pi\omega}.
\end{eqnarray*}
This kernel is no longer positive definite, but the solutions of the eigenvalue problem are still countable. This can easily be understood by deriving an equivalent kernel. First of all, all eigenvalues are bounded between $-1$ and $+1$ due to Eq.~(\ref{eq:criterion_augmented}). Second, we can offset the eigenvalues with $+1$ by adding the Dirac distribution $\delta(\omega)$ to the kernel, which does not modify the eigenfunctions. This equivalent kernel is now positive definite, ensuring the countability of its solutions and therefore the solutions of the augmented Slepian design are also countable. 

\begin{figure}
\includegraphics[width = \linewidth]{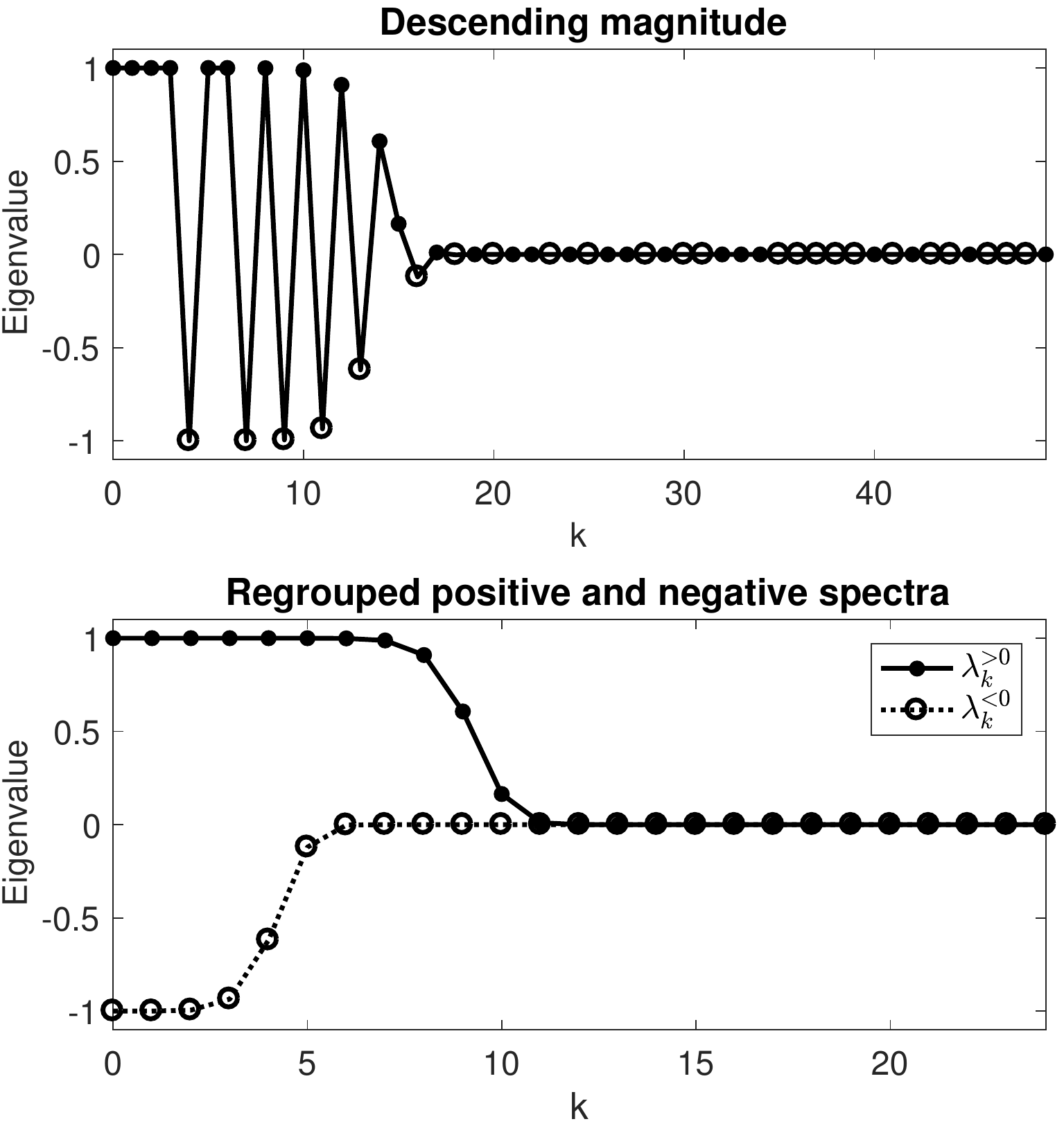}
\caption{Graphical representation of the splitting of the positive and negative parts of the spectrum. \label{fig:spectrum_def}}
\end{figure}

The eigenvalues cluster around three values: $+1$ for eigenfunctions well concentrated in $\mathcal{S}^+$, $-1$ for eigenfunctions well concentrated in $\mathcal{S}^-$, and $0$ for eigenfunctions that are neither concentrated in $\mathcal{S}^+$ nor $\mathcal{S}^-$. By convention, we will rank the eigenvalues $\lambda_k$, $k\in\mathbb{N}$, according to decreasing absolute value. We also introduce the following notation for positive $\lambda^{>0}_k$ and negative eigenvalues $\lambda^{<0}_k$, ranked according to decreasing \emph{absolute} value within their subsets. This grouping of eigenvalues is graphically represented in Fig.~\ref{fig:spectrum_def}.

We define the Shannon number $N_\text{Shannon}$ in the same way as in the original Slepian design, i.e. as the sum over all eigenvalues:
\begin{eqnarray}
N_\text{Shannon}&=&\sum_{k=0}^\infty\lambda_k\\
&=& \int_{-W}^W \lim_{\omega' \to \omega}D(\omega-\omega')d\omega\\ 
&=&\frac{2W\sum_{n=1}^{N^+}T^+_n}{\pi}-\frac{2W\sum_{n=1}^{N^-}T^-_n}{\pi}\label{eq:shannon_diff}\\
&=&\frac{2W\left(\sum_{n=1}^{N^+}T^+_n-\sum_{n=1}^{N^-}T^-_n\right)}{\pi}.
\end{eqnarray}
From Eq. (\ref{eq:shannon_diff}) and Eq. (\ref{eq:shannon}) it is straightforward to show that $N_\text{Shannon}$ is equal to the difference of the Shannon numbers $N^+_\text{Shannon}$ and $N^-_\text{Shannon}$ obtained for regular Slepians associated to $\mathcal{S}^+$ and $\mathcal{S}^-$, respectively: 
\begin{equation}
 N_\text{Shannon} = N_{\text{Shannon}}^+-N_{\text{Shannon}}^-.
 \label{eq:diff_shannon}
\end{equation}
We also use the notations $\lambda^+_k$ and $\lambda^-_k$ to refer to the eigenvalues of the regular Slepian constructions for $\mathcal{S}^+$ and $\mathcal{S}^-$, respectively.
From Eq. (\ref{eq:diff_shannon}), it is clear that $N_\text{Shannon}$ is equal to 0 if and only if the Shannon numbers of the regular Slepian constructions for both interval unions separately are equal. This can, for a given bandwidth, only happen if the total sizes of both unions are equal. As such, $N_\text{Shannon}$ is proportional to the size difference between both unions. 

\subsection{An instructive example}
\begin{figure}
\includegraphics[width = \linewidth, clip, trim={100 75 100 75}]{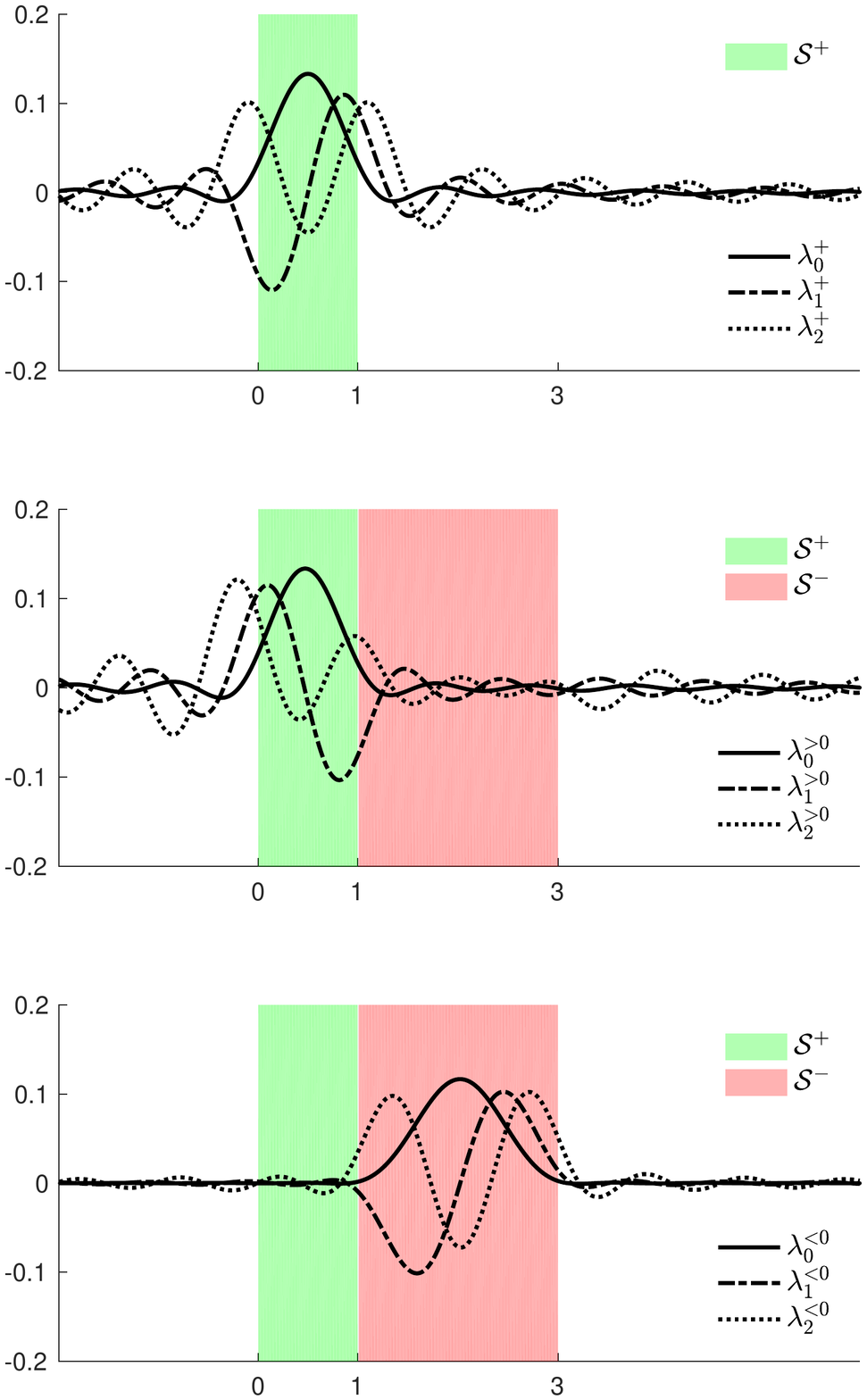}
\caption{Example of augmented Slepians, the one-sided continuous time bandwidth was set to $W = 2\pi$: (top) the first three Slepian functions resulting from the green selection are highly concentrated in the selected interval and are all even or odd around its center; (middle) the first three augmented Slepian functions with positive eigenvalues are still highly concentrated in the green interval, but are less concentrated in the red interval than the original Slepian functions; (bottom) the first three augmented Slepian functions with negative eigenvalues are highly concentrated in the red interval.\label{fig:slepian-1d}}
\end{figure}

In order to get a handle on what kind of results can be expected from the augmented Slepian framework, Fig.~\ref{fig:slepian-1d} shows an example where both intervals are put next to each other. All simulations in this and the following sections are based on the numerical method presented in Appendix \ref{app:simulation}. In the top figure, the first three original Slepian functions resulting from the selection of the green interval only are shown. These functions are highly concentrated in the green interval (i.e., $\lambda^+_k$ close to 1) and are always even or odd around the center of the interval.

The middle figure shows the first three augmented Slepian functions corresponding to positive eigenvalues when the red interval is negatively selected. It is clear that, while keeping high concentration within the green interval, these functions are not even or odd anymore around any point. Indeed, energy concentration is pushed away from the red interval and, therefore, more energy is located on the left side of the green interval. 

Finally, in the bottom figure, the first three augmented Slepian functions with negative eigenvalues are shown. As expected, they are highly concentrated within the red interval. Here, asymmetry exists as well since the green interval is now repulsing signal energy.

\subsection{Properties}
The classical PSWFs are known for a number of remarkably elegant properties. We now present how these original properties hold for augmented Slepians, as well as properties which are specific to the augmented setting. 

\begin{property}[Equivalence with conventional Slepian functions]
Augmented Slepians associated with $\mathcal{S}^-=\mathbb{R}\backslash \mathcal{S}^+$ are equivalent to conventional Slepian functions with $\mathcal{S}=\mathcal{S}^+$. 
\end{property}
\begin{IEEEproof}
It is straightforward to show that conventional Slepians are a special case. We plug $\mathcal{S}^-=\mathbb{R}\backslash \mathcal{S}^+$ into the energy optimization criterion (\ref{eq:criterion_augmented}) for augmented Slepians: 
\begin{eqnarray*}
  g(t) &=& \argmax_{g(t)\in\mathcal{B}_W}\left(\frac{ \int_{\mathcal{S}^+} |g(t)|^2 dt - \int_{\mathcal{S}^-} |g(t)|^2 dt}{ \int_\mathbb{R} |g(t)|^2 dt}\right)\label{eq:spec_1}\\
  &=&\argmax_{g(t)\in\mathcal{B}_W}\left(\frac{ \int_{\mathcal{S}^+} |g(t)|^2 dt - \int_{\mathbb{R}\backslash \mathcal{S}^+} |g(t)|^2 dt}{ \int_\mathbb{R} |g(t)|^2 dt}\right)\label{eq:spec_2}\\
  &=&\argmax_{g(t)\in\mathcal{B}_W}\left(\frac{ 2\int_{\mathcal{S}^+} |g(t)|^2 dt - \int_\mathbb{R} |g(t)|^2 dt}{ \int_\mathbb{R} |g(t)|^2 dt}\right)\label{eq:spec_3}\\
  &=&\argmax_{g(t)\in\mathcal{B}_W}\left(2\frac{ \int_{\mathcal{S}^+} |g(t)|^2 dt}{ \int_\mathbb{R} |g(t)|^2 dt}-1\right)\label{eq:spec_4}\\
  &=&\argmax_{g(t)\in\mathcal{B}_W}\left(\frac{ \int_{\mathcal{S}^+} |g(t)|^2 dt}{ \int_\mathbb{R} |g(t)|^2 dt}\right), \label{eq:spec_5}
\end{eqnarray*}
which reverts to the conventional criterion for $\mathcal{S}=\mathcal{S}^+$. 
\end{IEEEproof}

\begin{property}[Symmetry of solutions]
Interchanging the role of the union of intervals $\mathcal{S}^+$ and $\mathcal{S}^-$ as positive and negative selections in the design of augmented Slepians, leads to an equivalent solution where the signs of the eigenvalues $\lambda_k$ are inversed, but where the same associated eigenfunctions are found.
\end{property}

\begin{property}[Orthogonality] \label{prop:proof_ortho}
The augmented Slepians are double orthogonal over $\mathbb{R}$ and the union of intervals $\mathcal{S}^+$ and $\mathcal{S}^-$ in a generalized way: 
\begin{eqnarray}
  \left< g_k, g_l\right>_\mathbb{R} &=& \delta_{k-l}, \label{eq:fullortho}\\
  \left< g_k, g_l\right>_{\mathcal{S}^+} - \left< g_k, g_l\right>_{\mathcal{S}^-} &= & \lambda_k \delta_{k-l}.  \label{eq:partortho}
\end{eqnarray}
In addition, the augmented Slepians are approximately orthogonal on the union of intervals $\mathcal{S}^+$:
\begin{eqnarray}
k\ne l: & &\left| \left< g_k, g_l \right>_{\mathcal{S}^+} \right| \le \frac{\sqrt{(1-\lambda_k)(1-\lambda_l)}}{2}\\
k=l: & & \left<g_k, g_k \right>_{\mathcal{S}^+} \ge \lambda_k,
\end{eqnarray}
which become tight upper and lower bounds for Slepians well concentrated on $\mathcal{S}^+$ (i.e., $\lambda_k$ close to $1$). 
Notice that for $k\ne l$, we have $\left< g_k, g_l\right>_{\mathcal{S}^+}=\left<g_k,g_l\right>_{\mathcal{S}^-}$ due to Eq.~(\ref{eq:partortho}), and thus the orthogonality of these Slepians well concentrated on $\mathcal{S}^+$, becomes also strong on $\mathcal{S}^-$.
Similar results hold for the union of intervals $\mathcal{S}^-$: 
\begin{eqnarray}
k\ne l: & &\left| \left< g_k, g_l \right>_{\mathcal{S}^-} \right| \le \frac{\sqrt{(1+\lambda_k)(1+\lambda_l)}}{2}\\
k=l: & & \left<g_k, g_k \right>_{\mathcal{S}^-} \ge -\lambda_k,
\end{eqnarray}
which become tight for Slepians well concentrated on $\mathcal{S}^-$ (i.e., $\lambda_k$ close to $-1$).\\
Moreover, using the definition for the angle $\alpha$ between two eigenfunctions: 
\begin{eqnarray}
\cos\alpha_{g_k,g_l} = \dfrac{\left<g_k,g_l\right>}{||g_k||\cdot||g_l||},
\end{eqnarray}
we can show that the following property holds regarding the cosine of the angle between two eigenfunctions for $k\ne l$ and $\lambda_k>0$ and $\lambda_l>0$: 
\begin{eqnarray}
 \left|\cos \alpha_{g_k,g_l} \right| & \le & \frac{1}{2} \dfrac{\sqrt{ 1-\lambda_k } \sqrt{ 1-\lambda_l }}{\sqrt{\lambda_k\lambda_l}}.
\end{eqnarray}
For eigenfunctions for which $\lambda_k<0$ and $\lambda_l<0$ , the analogous property is given by:
\begin{eqnarray}
 \left|\cos \alpha_{g_k,g_l} \right| & \le & \frac{1}{2} \dfrac{\sqrt{ 1+\lambda_k } \sqrt{ 1+\lambda_l }}{\sqrt{\lambda_k\lambda_l}}.
\end{eqnarray}
\end{property}
The proof is given in Appendix~\ref{app:proof_ortho}.
 
\begin{figure*}[t]
\centering
\includegraphics[width = \linewidth, clip, trim={0 15 0 0}]{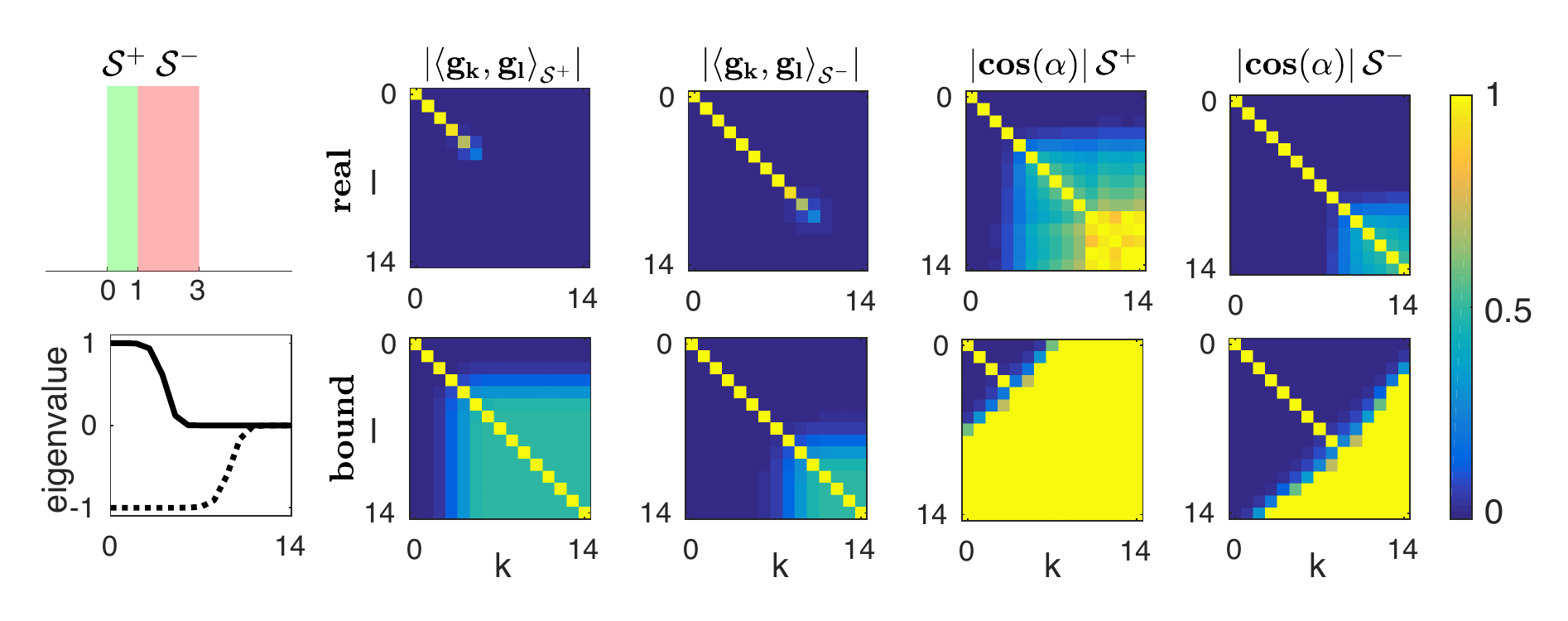}
\caption{The augmented Slepians are approximately orthogonal on the subsets $\mathcal{S}^+$ and $\mathcal{S}^-$, respectively. Comparison between the actual measures (top) and their bounds (bottom): inner product (left) and cosine angle (right), respectively. The selected intervals are shown on the top left and the one-sided bandwidth is 5\% of the Nyquist frequency.} 
\label{fig:approx_orthogonality}
\end{figure*}

In Fig.~\ref{fig:approx_orthogonality}, we show an example comparing the actual inner products and the cosines together with their respective bounds. 

\begin{property}[Interaction parameter\label{prop:interaction}] 
We introduce the interaction parameter $\Delta^+$ as the difference between the Shannon number of the conventional Slepian design for $\mathcal{S^+}$ and the sum of the positive eigenvalues of the augmented Slepian spectrum. It turns out to be equal to the interaction $\Delta^-$ between the conventional design for $\mathcal{S^-}$ and the negative part of the augmented Slepian spectrum: 
\begin{equation}
\Delta = \underbrace{N_{Shannon}^+-\sum^\infty_{k=0}\lambda^{>0}_{k}}_{=\Delta^+} = \underbrace{N_{Shannon}^-+\sum^\infty_{k=0}\lambda^{<0}_{k}}_{=\Delta^-}.
\end{equation}
\end{property}
The proof is given in Appendix~\ref{app:interaction}.

\begin{figure}
\centering
\includegraphics[width = .95\linewidth,clip,trim={0.5 0 0 0}]{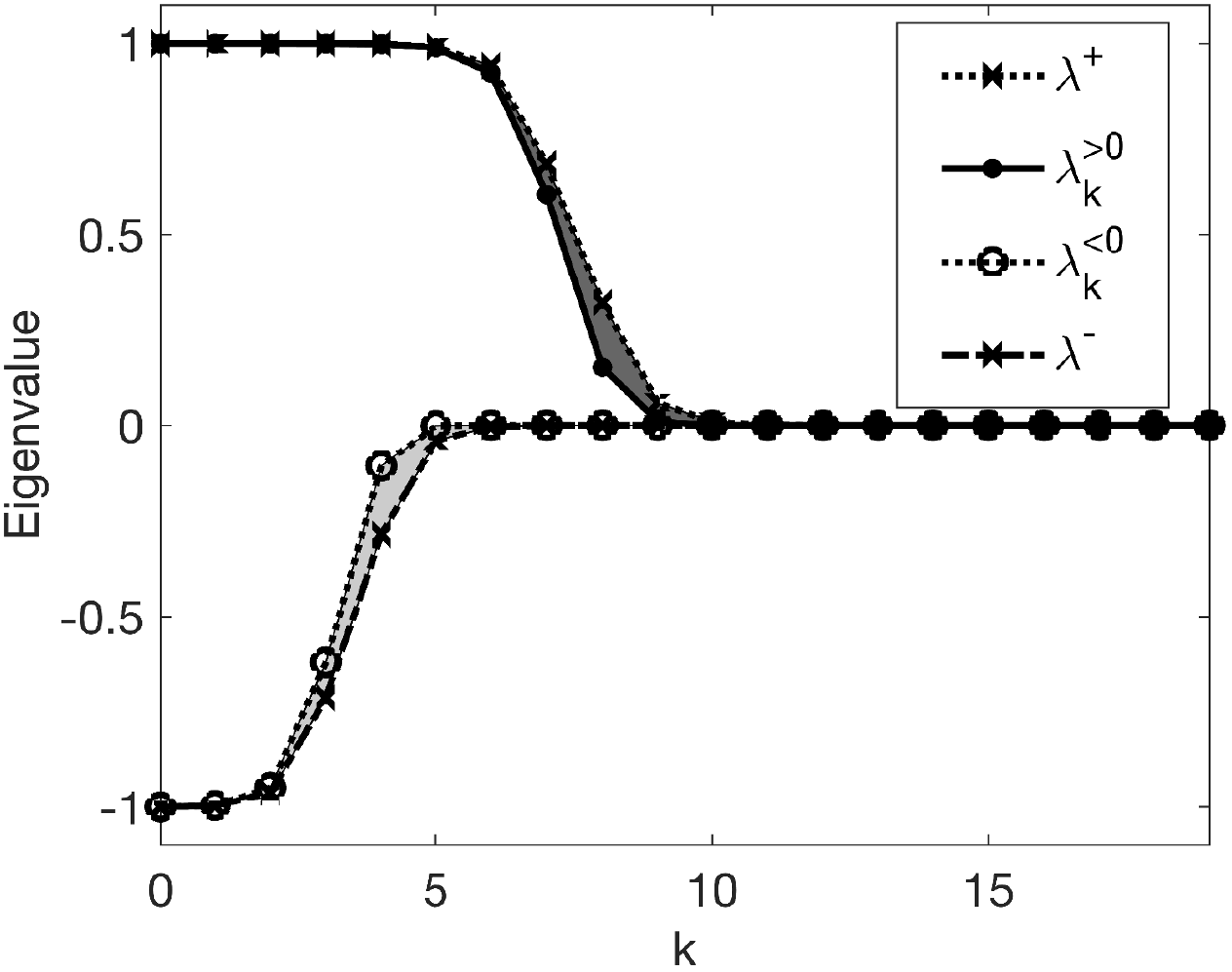}
\caption{Eigenvalue spectra for augmented and regular Slepian designs on the positively and negatively selected intervals, respectively. The interaction parameter $\Delta$ corresponds to each of the shaded surface areas.\label{fig:sketch_deltas}
}
\end{figure}

A visual interpretation of these parameters is shown in Fig.~\ref{fig:sketch_deltas}. These values can be used to quantify how adding a negative region to the concentration problem makes it more difficult to achieve high (generalized) energy concentration in the original interval when this region is placed close to the positive region. However, a wide spacing of the regions does not highly influence the achievable energy concentration in the positively selected region. 

\section{Discussion\label{sec:Discussion}}


Now that the theoretical framework and properties of the augmented Slepians have been introduced, we will discuss in more details some of their features, including their importance for practical applications.


\subsection{Interplay between two types of intervals}
The main advantage of the proposed design is the possibility to specify two types of intervals that play a different role in the optimization criterion. Consequently, while a single basis is obtained, Slepian functions that are well-localized in one versus the other type of interval are associated with different eigenvalues; i.e., positive and negative ones, respectively. One question then is whether a similar result could have been obtained by combining two conventional Slepian bases. The answer is no because such a dual construction would not have led to orthogonality properties on the intervals. In particular, as shown by Property~\ref{prop:proof_ortho}, Slepians well concentrated on $\mathcal{S}^+$ (i.e., $\lambda_k$ close to $1$) are (approximately) orthogonal on both $\mathcal{S}^+$ and $\mathcal{S}^-$ taken separately. Consequently, inner products taken on either of the different intervals between a signal and the augmented Slepians can be considered independent.

\begin{figure}[h]
    \centering 
    \includegraphics[width=\linewidth, clip, trim={0.5 0 0 0}]{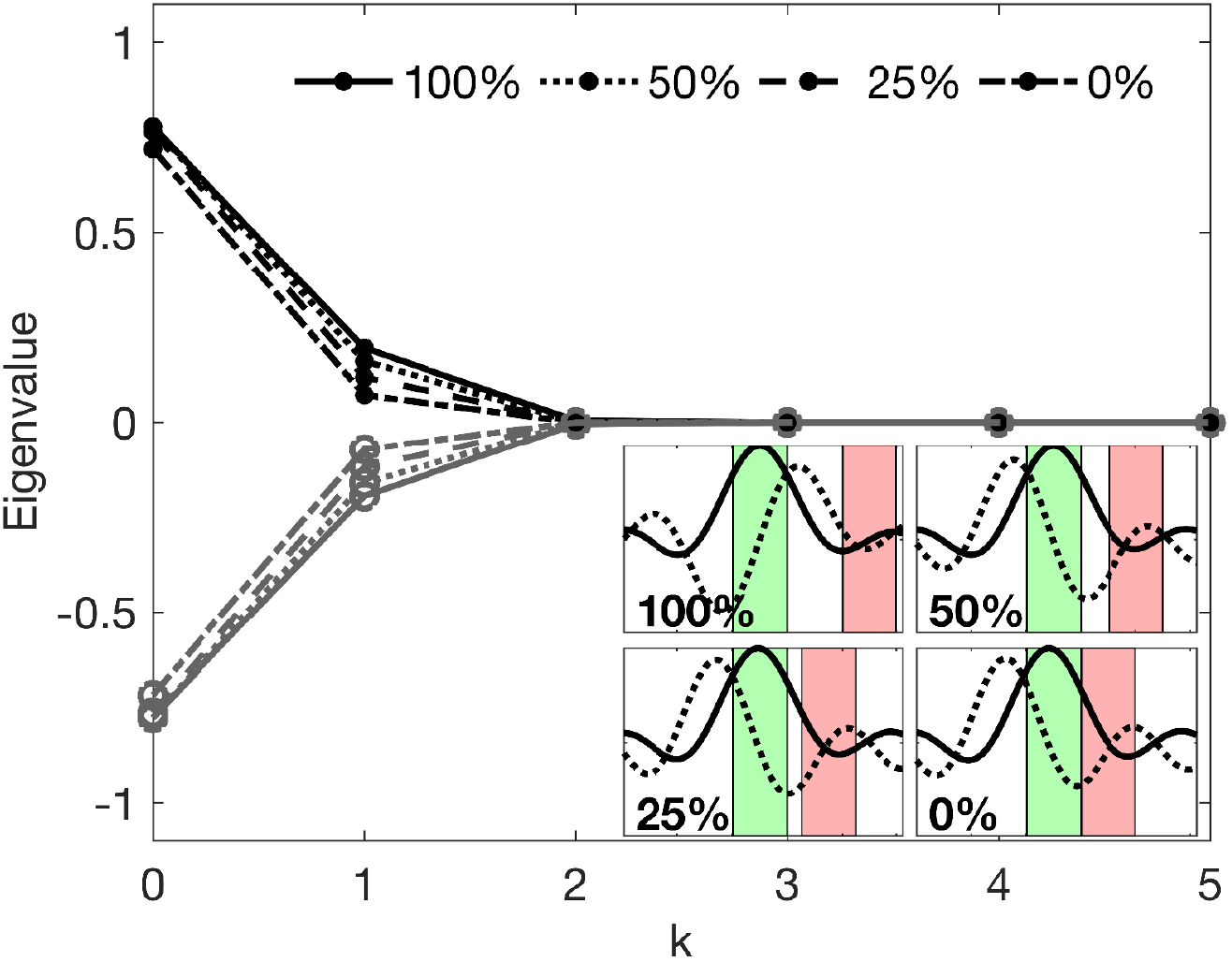}
    \caption{Eigenvalue spectrum of augmented Slepian design for different spacings between two equally sized intervals that are positively and negatively weighted, respectively. The distance between the two intervals is changed and specified as the percentage of the interval size. The insets show the corresponding first two eigenfunctions associated with the two largest eigenvalues.
    \label{fig:interaction}
    }
 \end{figure}

\begin{figure}
\centering
\includegraphics[width=\linewidth]{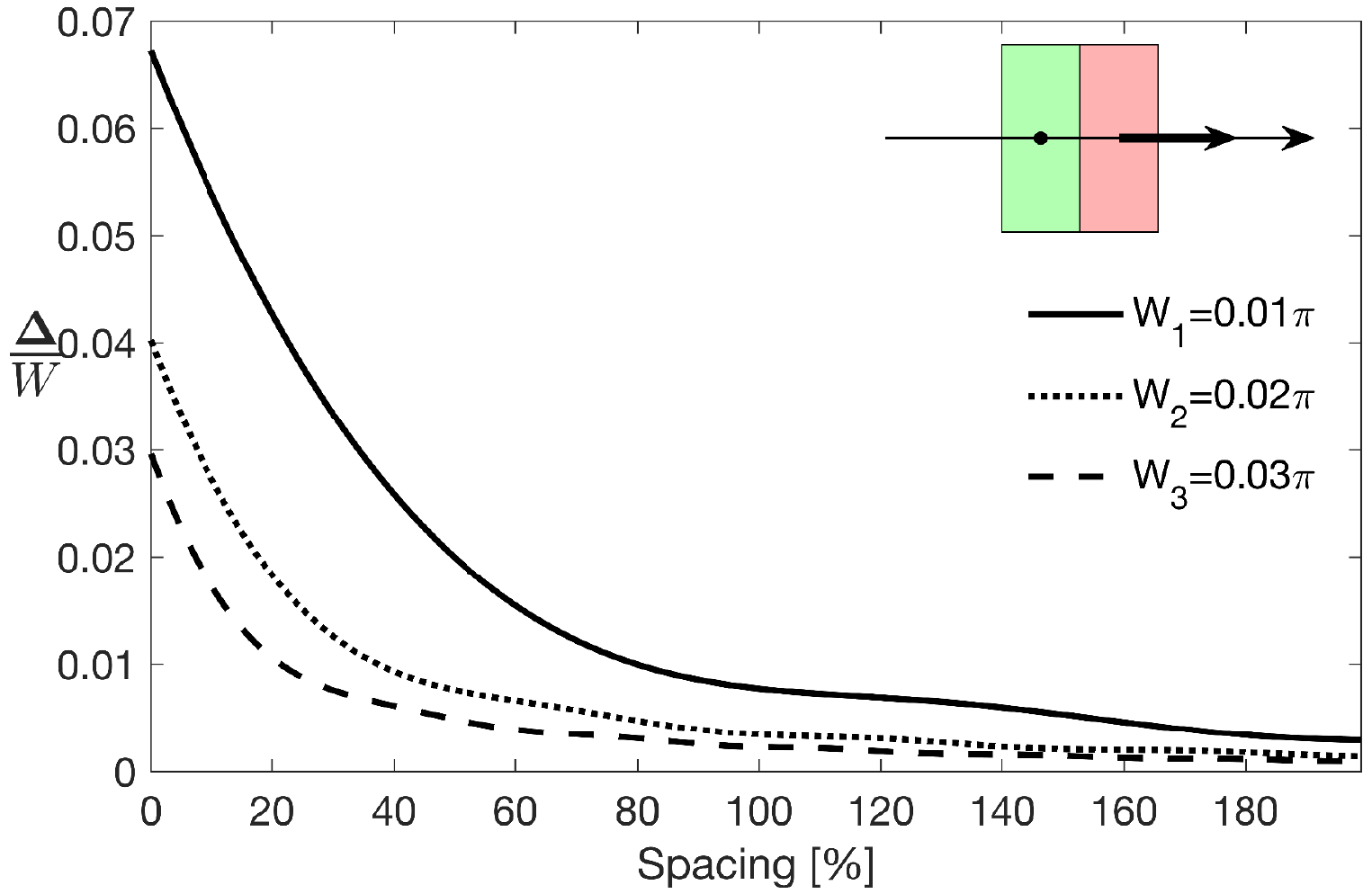}
\caption{Interaction parameter $\Delta$ normalised by the bandwidth $W$ as a function of the spacing between selected and penalised intervals for 3 different bandwidths. As expected, the interaction parameter decreases as the intervals are spaced more largely. Increasing bandwidth lowers the normalised interaction parameter, but the behavior as function of spacing remains similar.
\label{fig:interaction_decay}}
\end{figure}

In addition to the orthogonality and independence properties, when the two types of intervals are close enough, an effect of interaction can be observed on the eigenvalue spectrum as quantified by $\Delta$ of Property \ref{prop:interaction}. This parameter can be interpreted as the difference in energy concentration between conventional Slepians for $\mathcal{S}^+$, and the augmented Slepians that come with positive eigenvalues and thus are more concentrated in $\mathcal{S}^+$ than $\mathcal{S}^-$.  This phenomenon is illustrated in Figs.~\ref{fig:interaction} and \ref{fig:interaction_decay}. In particular, Fig.~\ref{fig:interaction} shows how bringing the intervals closer together shifts the positive (resp. negative) part of the spectrum downward (resp. upward). Also the first two augmented Slepian functions are shown on the insets; in (a), the functions resemble more conventional Slepians (i.e., even and odd with respect to the center of the green interval, so no preference for a certain side), in (b)-(d), the (anti-)symmetry gets lost as the intervals move closer together which is indicative for the interaction. Fig.~\ref{fig:interaction_decay} shows the evolution of the interaction parameter $\Delta$ normalized by the bandwidth as a function of the spacing between the two intervals of interest. As expected, $\Delta$ decreases with increased spacing. The results for 3 different bandwidths are shown and, although the relationship between $\Delta$ and $W$ is clearly not linear since the curves do not coincide, they roughly have the same shape. On a side note, if the bandwidth is infinite, the spectrum will always show a perfect step-like shape in both the conventional and the augmented Slepian frameworks irrespectively of the spacing between the regions of interest. Therefore, in this extreme case, $\Delta$ will always be equal to $0$.

\begin{figure}
\includegraphics[width =\linewidth,clip,trim={80 160 100 160}]{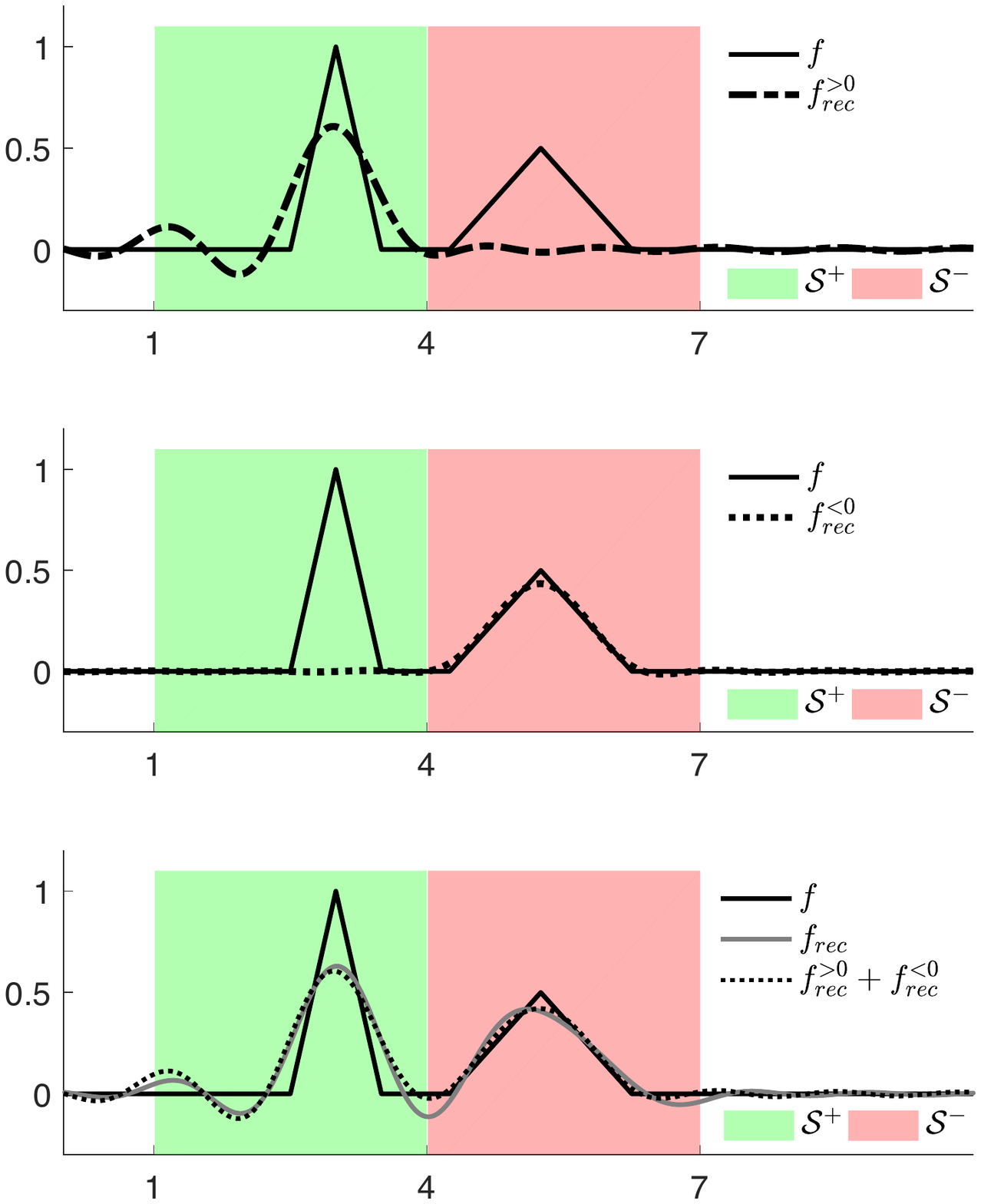}
\caption{Bandlimited reconstruction of a function in two intervals independently. The original signal is shown together with the reconstructions based on the augmented Slepian design and on the original Slepian construction for the union of both intervals. \label{fig:reconstruction_example}}
\end{figure}

The clear distinction in the eigenvalue spectrum between functions having high concentration in either of the intervals can be exploited when considering the bandlimited reconstruction of a signal on distinct intervals, for instance. We provide an example of how the augmented Slepian design can be used to reconstruct a signal on two intervals independently, though being linked by the same decomposition. We start from measurements in the Fourier domain, which is suggestive of measurements taken in $k$-space as in magnetic resonance imaging. 
Let us consider the signal $f(t)$ shown by the black full line in Figure \ref{fig:reconstruction_example}. 
Assume now that we want to reconstruct the signal in the intervals $\mathcal{S}^+=[1,4]$ and $\mathcal{S}^-=[4,7]$, independently, and that data can only be acquired within the bandwidth $[-W,W]$ where $W=1.5\pi$. We can then use the augmented Slepian design with band-limit $W$ to reconstruct the signals as follows:
\begin{eqnarray}
f^{>0}_\text{rec}(t) &=& \sum_{n=0}^{N^+_{\text{Shannon}}}\left<G^{>0}_n, F\right>_\mathbb{R}g^{>0}_n(t)\\
f^{<0}_\text{rec}(t) &=& \sum_{n=0}^{N^-_{\text{Shannon}}}\left<G^{<0}_n, F\right>_\mathbb{R}g^{<0}_n(t),
\end{eqnarray}
where $F$ is the Fourier transform of $f$. 
Figure \ref{fig:reconstruction_example} shows the original signal and the reconstructed signals on both intervals, as well as the reconstructed signal $f_\text{rec}$ using the original Slepian design on the union of $\mathcal{S}^+$ and $\mathcal{S}^-$ with the total number of eigenfunctions used for the augmented design (i.e., $N_{\text{Shannon}} = N^+_{\text{Shannon}}+N^-_{\text{Shannon}}$). The reconstructions $f_\text{rec}^{>0}$ and $f_\text{rec}^{<0}$ are well localized within their respective intervals. By construction, these reconstructions are orthogonal and thus explain separate parts of the measured energy. The sum of both reconstructions is very close to the one using conventional Slepians on the combined intervals, but it approximates better the ground truth at the boundary between $\mathcal{S}^+$ and $\mathcal{S}^-$, which is unknown to the conventional design. This example illustrates how the proposed design can be beneficial when additional prior information is available to tailor band-limited reconstructions.

\subsection{Alternative way to penalize energy concentration}
In less known work, Gilbert and Slepian~\cite{Gilbert.1977} have proposed a generalization of the Slepian functions that maximizes the ratio
\begin{equation}
  \lambda = \max_{g(t)\in {\mathcal B}_W}\left(\frac{ \int_\mathcal{S^+} \left|g(t)\right|^2 dt}{ \int_{\mathcal{S}^-} \left|g(t)\right|^2 dt}\right),
\end{equation}
which reverts to the original Slepian design in case $\mathcal{S}^- = \mathbb{R}$. In this theory, the ratio between the energy concentrations in $\mathcal{S}^+$ and $\mathcal{S}^-$ is optimized as opposed to their difference in our augmented Slepian design.

Although the resulting functions have the nice property of being orthogonal on both intervals separately, the eigenvalue spectrum does in general not show the striking phase transitions visible with the augmented Slepian framework. 

Moreover, Gilbert and Slepian reported that a corresponding differential equation (i.e., the lucky accident~\cite{Slepian.1983}) could only be found for special cases such as in the original Slepian design. The reason behind is that the differential operator needs to commute with the characteristic function~\cite{Grunbaum.1982,Walter.1992}. Therefore, it might not be possible to find such operator for the augmented Slepian design either, except in some particular choices of the intervals.

\subsection{Applications and extensions}
Given its fundamental, but at the same time practical objectives, the original Slepian framework has found a wide range of applications, ranging from signal processing (filtering and multitaper spectral analysis~\cite{Mathews.1985}, extrapolation and compressed sensing~\cite{Gosse.2010,Davenport.2012,Gosse2013}, compression~\cite{Karnik2016}) to geophysics~\cite{Simons.2000,Albertella.2000,Simons.2005}, ultrawideband communications (to describe radiation patterns of antennas~\cite{Dullaert.2010} or pulse designs~\cite{Wu.2013,Anderson.2017}), and magnetic resonance imaging (for extrapolation~\cite{Plevritis.1995}, speeding up data acquisition within a predefined region-of-interest~\cite{Xang.2002}). Many of these applications have been built upon extensions of the original framework to higher-dimensional spaces~\cite{Slepian.1964}, to the sphere~\cite{Miranian.2004,Simons_2006,Bates2016}, or more recently to graphs~\cite{Tsitsvero2016,vandeville1701,VanDeVille2017}. Other generalizations have been proposed for a weighted criterion to optimize steerable filters~\cite{unser1301}, for the quaternionic Fourier transform~ \cite{Zou2016}, or for matrix-valued functions \cite{grunbaum2015}. Even a Fast Slepian Transform has been proposed as an alternative to the Fast Fourier Transform for time-limited signals \cite{Karnik2017}.

The proposed design of augmented Slepians can probably be made useful in many of these applications. In extrapolation, for instance, conventional Slepians are used to compute inner products on an interval where measures are available, to then be used to obtain a bandlimited extrapolation. With the proposed design, two separated intervals could be specified and lead to two extrapolations, but that would remain orthogonal thanks to the joint optimization criterion. However, extrapolation requires the calculation of Slepian functions corresponding to eigenvalues close to zero~\cite{Gosse.2010}. At some point, these functions and eigenvalues cannot be calculated precisely enough using the numerical procedure presented in Appendix \ref{app:simulation} to provide good extrapolation results. For the original Slepian design, the commuting differential equation (or ``lucky accident'') has been used to provide an alternative way to calculate the eigenfunctions more precisely. Further research into more accurate calculation methods, and whether a commuting kernel exists, could be extremely useful to pave the way for application of the augmented Slepian framework.

Extending the augmented Slepian design to more-dimensional and more complex spaces should also be possible. For this, writing the optimisation problem using an operator formalism can be instructive:
\begin{eqnarray}
\mathcal{B}(\mathcal{D}^{\mathcal{S}^+}-\mathcal{D}^{\mathcal{S}^-})\mathcal{B}g = \lambda g,
\end{eqnarray}
where $\mathcal{B}$ is the band-limiting operator and $\mathcal{D}^{(\cdot)}$ the time-limiting operator.

\subsection{Indefinite inner product and Krein spaces}
There is an interesting link between the kernel $D$ of the augmented Slepian design and Krein spaces~\cite{Azizov.1989}. 
In fact, since $D$ is indefinite (i.e., it has both positive and negative eigenvalues), we cannot define a Hilbert space based on it, but it is possible to define a Krein space $\mathcal{K}$, by building on the indefinite inner product over the generalized selection $\mathcal{S}^+ \cup \mathcal{S}^-$:
\begin{equation}
 \left( x,y \right) = \underbrace{\int_{\mathcal{S}^+}x(t)\bar{y}(t)dt}_{\left< x, y \right>_{\mathcal{S}^+}} \underbrace{-  \int_{\mathcal{S}^-}x(t)\bar{y}(t)dt}_{-\left< x,y \right>_{\mathcal{S}^-}},
\end{equation}
which admits a direct orthogonal sum decomposition $$\mathcal{K}=\mathcal{K}^+ \oplus \mathcal{K}^-,$$ where $\mathcal{K}^+, \left< \cdot, \cdot \right>_{\mathcal{S}^+}$ and $\mathcal{K}^-, -\left<\cdot,\cdot\right>_{\mathcal{S}^-}$ are Hilbert spaces, and which has $\left(x, y \right)=0$ for any $x\in \mathcal{K}^+$, $y\in \mathcal{K}^-$. 

This also means that we can define the projection operators that map onto these constituent spaces as $\mathcal{K}^+=\mathcal{P}^+ \mathcal{K}$ and $\mathcal{K}^-=\mathcal{P}^- \mathcal{K}$, which can then be combined in an endomorphism operator on $\mathcal{K}$ as $\mathcal{J} = \mathcal{P}^+ - \mathcal{P}^-$; i.e., this operator defines a positive semi-definite inner product:
 \begin{eqnarray}
 \left< x,y \right> &\stackrel{\text{def}}{=}& (x, \mathcal{J}y)\\
 &=&(x, (\mathcal{P}^+ - \mathcal{P}^-)y )\\
 &=& \int_{\mathcal{S}^+}x(t)\bar{y}(t)dt +  \int_{\mathcal{S}^-}x(t)\bar{y}(t)dt\\
 & =&  \int_{\mathcal{S}^+\cup\mathcal{S}^-}x(t)\bar{y}(t)dt 
 \end{eqnarray}
 and satisfies the property $\mathcal{J}^3 = \mathcal{J}$. It can be readily verified that applying the kernel three times indeed reverts to a single application.

In many application fields such as data analysis and learning tasks, kernels are typically required to be positive semi-definite, however, there is also an interest in using non-positive kernels~\cite{Ong.2004,Liwicki.2012} and therefore the augmented Slepians might be useful to guide new designs in this much larger search space.

\subsection{Generalized weightings}
The energy concentrations within the intervals are subtracted and thus weighted with coefficients -1 and 1, respectively. Obviously these weights could be changed depending on the application; e.g., reciprocally scaled w.r.t. the size of the intervals. 
One might consider this as a particular case of expressing the energy concentration using weighting functions that vary within each interval; i.e., the criterion could be generalized as 
\begin{equation}
  \lambda = \max_{g(t)\in {\mathcal B}_W}\left(\frac{ \int_\mathcal{S^+} w^+(t)\left|g(t)\right|^2 dt -  \int_\mathcal{S^-}  w^-(t)\left|g(t)\right|^2 dt}{ \int_\mathbb{R} \left|g(t)\right|^2 dt}\right),
\end{equation}
where $w^+(t)$ and $w^-(t)$ are positive real-valued functions.
Turning this into the Fourier domain, we get the following eigenvalue equation:
\begin{eqnarray}
 \int_{-W}^W D(\omega-\omega^\prime)G(\omega^\prime)d\omega^\prime&=& \lambda G(\omega),
\end{eqnarray}
where
\begin{eqnarray*}
D(\omega) & = & W^+(\omega) - W^-(\omega)
\end{eqnarray*}
with $W^+$ and $W^-$ the Fourier transformed window functions.
As with the original and augmented Slepian framework, this equation is still a homogeneous Fredholm equation of the second kind with a Hermitian kernel.

This more general optimisation problem is a special case, as is the original Slepian framework, of Franks' variational framework where a general optimisation criterion is built with energy constraints in both time and Fourier domain \cite{Franks_1969}. It can also be seen as a special case of the pseudo-differential operator framework and the corresponding asymptotic theory presented in \cite{Sirovich1985,Victor2003}, which can be useful for further theoretical study and extension of the concept of augmented Slepians.
Most properties outlined here are tightly linked to the specific case of having weights -1 and 1, and, therefore, further research is needed to better understand how these properties can be further generalised for more general weighting functions. 

A final question that emerges naturally about the extension presented here, is whether or not it is possible to consider more than two types of intervals. Intriguingly, such goal can be reached when the weighting factors are allowed to be complex-valued. For instance, we can select weights of 1, $\exp(i\frac{2\pi}{3})$ and $\exp(-i\frac{2\pi}{3})$ for three different types of intervals, which provides an eigenvalue spectrum where eigenvalues are closely located to lines in the complex plane with angles 0, $\frac{2\pi}{3}$ and $\frac{-2\pi}{3}$, respectively.

\section{Conclusion}
We presented an extension of the Slepian design that leads to band-limited functions that simultaneously maximize and minimize energy concentration in different types of intervals. We showed the mathematical background of these ``augmented'' Slepian functions, together with their main properties and how they can be practically obtained. The eigenvalue spectrum exhibits some essential features such as two phase transitions---one for each type of interval. The degree of ``interaction'' between both intervals is also embedded in the eigenvalue spectrum. Just as regular Slepian functions, their augmented variants are orthogonal over the whole domain, in a generalized way over the selected intervals, and approximately (within given bounds) over the intervals of each type. Given the broad impact of Slepian functions, we expect this work can find various applications. 
\appendices

\section{Proof of Proposition \ref{prop:translated_temporal_interval}\label{app:translated_temporal_interval}}
\begin{IEEEproof}
Using the notation $\sigma=TW$, we postulate the following variant of the Fourier property for the case of a shifted interval
\begin{equation}
   \psi_k(t) = \frac{\myexp{i P W t}}{\mu_k} \int_{-1}^{+1} \psi_k(\xi) \myexp{i (PW+\sigma t)\xi} d\xi,
\end{equation}
into which we plug the complex conjugate
\begin{equation}
   \psi_k(\xi) = \frac{\myexp{-i P W\xi}}{\bar{\mu}_k} \int_{-1}^{+1} \psi_k(\xi') \myexp{-i(PW+\sigma\xi)\xi'} d\xi'.
\end{equation}
which leads to
\begin{equation}
   \psi_k(t) = \frac{\myexp{iPWt}}{\left|\mu_k\right|^2} \int_{-1}^{+1} \psi_k(\xi') \myexp{-iPW\xi'} \int_{-1}^{+1}  \myexp{i \sigma (t-\xi') \xi} d\xi d\xi'.
\end{equation}
With the change of variable $w=\sigma\xi$, we obtain:
\begin{equation*}
   \psi_k(t) = \frac{\myexp{iPWt}}{\left|\mu_k\right|^2\sigma} \int_{-1}^{+1} \psi_k(\xi') \myexp{-i PW\xi'} \int_{-\sigma}^{+\sigma}  \myexp{i (t-\xi') w} dw d\xi'.
\end{equation*}
Using the inverse Fourier transform of the window function $[-\sigma,+\sigma]$, we further obtain
\begin{equation}
   \psi_k(t) = \frac{2\pi}{\left|\mu_k\right|^2 \sigma} \int_{-1}^{+1} \psi_k(\xi') \underbrace{\myexp{iPW(t-\xi')} \frac{\sin(\sigma(t-\xi'))}{\pi(t-\xi')}}_{=D(t-\xi')} d\xi',
\end{equation}
which is the integral equation that we would obtain by expressing the maximal energy concentration in the shifted interval; i.e., the kernel $D$ can be directly related to its Fourier transform. In addition, we identified the relationship $\lambda_k=\left|\mu_k\right|^2 \sigma / (2\pi)$, which is the same as for the conventional PSWF.
\end{IEEEproof}

\section{Proof of Proposition \ref{prop:ortho}\label{app:ortho}}
\begin{IEEEproof}
The first property is trivial given that the Slepian functions $g_k$ are eigenfunctions and thus orthogonal and normalized such that $\left\| g_k \right\|^2=1$. The second property can be derived as follows:
\begin{eqnarray*}
\left< g_k,g_l \right>_\mathcal{S} & = & \int_\mathcal{S} g_k(t) \bar{g}_l(t) dt \\
& = & \int_\mathcal{S} g_k(t) \frac{1}{2\pi} \int_{-W}^{+W} \bar{G}_l(\omega) \myexp{-i\omega t} d\omega dt \\
& = & \frac{1}{2\pi} \int_{-W}^{+W} \bar{G}_l(\omega) \int_\mathcal{S} g_k(t) \myexp{-i\omega t} dt d\omega \\
& = & \frac{1}{2\pi} \int_{-W}^{+W} \bar{G}_l(\omega) \int_{-W}^{+W} G_k(\omega') D(\omega'-\omega) d\omega' d\omega \\
& = & \frac{1}{2\pi} \int_{-W}^{+W} G_k(\omega') \int_{-W}^{+W} \bar{G}_l(\omega) D(\omega'-\omega) d\omega d\omega' \\
& = & \frac{\lambda_l}{2\pi} \int_{-W}^{+W} G_k(\omega) \bar{G}_l(\omega) d\omega = \lambda_l \delta_{k-l}.
\end{eqnarray*}
\end{IEEEproof}

\section{Proof of Property \ref{prop:proof_ortho}\label{app:proof_ortho}}
\begin{IEEEproof} We start from the following energy-concentration property which follows from the definition of the augmented Slepian functions:
\begin{eqnarray}
  \left< g_k, g_k \right>_{\mathcal{S}^+} - \left< g_k, g_k \right>_{\mathcal{S}^-} &=& \lambda_k \left< g_k, g_k \right>.
\end{eqnarray}
Since $\left< g_k, g_k \right>_{\mathcal{S}^-}\ge 0$ and $\left< g_k, g_k \right>=1$, this means that
\begin{eqnarray}
  \label{eq:tmp1}
  \left< g_k, g_k \right>_{\mathcal{S}^+} &\ge& \lambda_k.
\end{eqnarray}
When $k\ne l$, Eq.~\ref{eq:fullortho} can be rewritten as the sum of its parts:
\begin{eqnarray}
  \left< g_k, g_l\right>_{\mathcal{S}^+} + \left< g_k, g_l\right>_{\mathcal{S}^-} + \left< g_k, g_l\right>_{\mathcal{S}^*}&=& 0,
\end{eqnarray}
where $\mathcal{S}^*$ is the full domain minus $\mathcal{S}^+$ and $\mathcal{S}^-$. Using Eq.~(\ref{eq:partortho}), it follows that 
\begin{eqnarray}
  \left< g_k, g_l\right>_{\mathcal{S}^+} = - \frac{1}{2} \left< g_k, g_l\right>_{\mathcal{S}^*}\label{eq:tmp2}.
\end{eqnarray}
Applying Cauchy-Schwartz to $\left< g_k, g_l\right>_{\mathcal{S}^*}$ then shows
\begin{eqnarray}
  \label{eq:cauchy}
  \left| \left< g_k, g_l\right>_{\mathcal{S}^*} \right| & \le & \sqrt{ \left< g_k, g_k\right>_{\mathcal{S}^*} } \sqrt{ \left< g_l, g_l\right>_{\mathcal{S}^*} },
\end{eqnarray}
where the right-hand terms can be rewritten as
\begin{eqnarray}
  \sqrt{ \left< g_k, g_k\right>_{\mathcal{S}^*} } &=& \sqrt{ \left< g_k, g_k\right> - \left< g_k, g_k\right>_{\mathcal{S}^+} - \left< g_k, g_k\right>_{\mathcal{S}^-}} \nonumber\\
  & \le & \sqrt{ 1 - \left< g_k, g_k\right>_{\mathcal{S}^+}} \nonumber\\
  & \le & \sqrt{ 1 - \lambda_k},\quad\text{using (\ref{eq:tmp1})}.
\end{eqnarray} 
This further simplifies Eq.~\ref{eq:cauchy} into
\begin{eqnarray}
  \left| \left< g_k, g_l\right>_{\mathcal{S}^*} \right| & \le & \sqrt{ 1-\lambda_k } \sqrt{ 1-\lambda_l }
\end{eqnarray}
and using this in Eq.~\ref{eq:tmp2}, the following bound can be found:
\begin{eqnarray}
  \left| \left< g_k, g_l\right>_{\mathcal{S}^+} \right| & \le & \frac{1}{2} \sqrt{ 1-\lambda_k } \sqrt{ 1-\lambda_l }\label{eq:initial_bound}.
\end{eqnarray}
In order to rule out the effect of the signal magnitude inside the region of interest, the geometrical definition of the inner product is used in Eq.~\ref{eq:initial_bound}:
\begin{eqnarray}
 |g_k|_{\mathcal{S}^+} |g_l|_{\mathcal{S}^+}|\cos \alpha_{g_k,g_l} | & \le & \frac{1}{2} \sqrt{ 1-\lambda_k } \sqrt{ 1-\lambda_l }.\label{eq:innerprodsubstituted}
\end{eqnarray}

Using $\lambda_k = |g_k|^2_{\mathcal{S}^+}-|g_k|^2_{\mathcal{S}^-}$, it can be concluded that $\lambda_k \le |g_k|^2_{S^+}$. For all $\lambda_k>0$ it is then also true that $\sqrt{\lambda_k}\le |g_k|_{\mathcal{S}^+}$ and since $\lambda_k \le 1$ and $|g_k|_{\mathcal{S}^+}\le1$ this leads to:
\begin{eqnarray}
\frac{1}{|g_k|^{}_{\mathcal{S}^+}}&\le\dfrac{1}{\sqrt{\lambda_k}}\quad\text{ if }\lambda_k > 0.
\end{eqnarray}
Using this inequality in Eq.~\ref{eq:innerprodsubstituted}, the final bound on the approximate orthogonality on the positively selected region of interest for the eigenfunctions for which $\lambda>0$ becomes:
\begin{eqnarray}
 |\cos \alpha_{g_k,g_l} | & \le & \frac{1}{2} \dfrac{\sqrt{ 1-\lambda_k } \sqrt{ 1-\lambda_l }}{\sqrt{\lambda_k\lambda_l}} \text{ if }\lambda_{k,l}>0
 \end{eqnarray}

A fully analogous derivation leads to a bound on the approximate orthogonality on the positively selected region of interest for the eigenfunctions for which $\lambda<0$ :
\begin{eqnarray}
 |\cos \alpha_{g_k,g_l} | & \le & \frac{1}{2} \dfrac{\sqrt{ 1+\lambda_k } \sqrt{ 1+\lambda_l }}{\sqrt{\lambda_k\lambda_l}} \text{ if }\lambda_{k,l}<0.
 \end{eqnarray}
 \end{IEEEproof}

\section{Proof of Property \ref{prop:interaction}\label{app:interaction}}
\begin{IEEEproof}[Proof]
Using the identity
\begin{eqnarray}
\sum^\infty_{k=0}\lambda_k &=&\sum^\infty_{k=0}\lambda^+_k - \sum^\infty_{k=0}\lambda^-_k,
\end{eqnarray}
which follows from Eq. \ref{eq:diff_shannon}, and the fact that $\sum^\infty_{k=0} \lambda_k$ can be written as the sum of its positive and negative parts: 
\begin{eqnarray}
\sum^\infty_{k=0} \lambda_k = \sum^\infty_{k=0} \lambda^{>0}_{k} + \sum^\infty_{k=0} \lambda^{<0}_{k},
\end{eqnarray}
the difference  $\Delta^+ - \Delta^-$ becomes
\begin{eqnarray}
&=&  \sum^\infty_{k=0}\lambda^+_k - \sum^\infty_{k=0}\lambda^{>0}_{k} - (\sum^\infty_{k=0}\lambda^-_k + \sum^\infty_{k=0}\lambda^{<0}_{k})\\
&=&\sum^\infty_{k=0}\lambda^+_k - \sum^\infty_{k=0}\lambda^-_k - \sum^\infty_{k=0}\lambda^{>0}_{k} - \sum^\infty_{k=0}\lambda^{<0}_{k}\\
&=& (\sum^\infty_{k=0}\lambda^+_k - \sum^\infty_{k=0}\lambda^-_k) - (\sum^\infty_{k=0}\lambda^{>0}_{k} + \sum^\infty_{k=0}\lambda^{<0}_{k})\\
&=&(\sum^\infty_{k=0}\lambda^+_k - \sum^\infty_{k=0}\lambda^-_k) - \sum^\infty_{k=0}\lambda_k\\
&=&\sum^\infty_{k=0}\lambda_k-\sum^\infty_{k=0}\lambda_k\\
&=&0
\end{eqnarray}
This finishes the proof that $\Delta^+$ and $\Delta^-$ are equal.
\end{IEEEproof}

\section{Numerical Method\label{app:simulation}}
While the theoretical developments in this work were in the continuous domain, all examples were simulated numerically and thus in the discrete domain. The original Slepian optimization criterion in discrete time can be written as a Rayleigh quotient:
\begin{eqnarray}
  \label{eq:slepian-1d-Rayleigh}
   \lambda = \frac{ \vc v^H \ma C \vc v}{\vc v^H \vc v},
\end{eqnarray}
where $\ma C = \ma F_W^H\ma S^+\ma F_W$ is the concentration matrix with $\ma F_W$ the unitary Discrete Fourier Transform matrix limited to the selected frequency band (bandwidth $W$) and $\ma S^+$ is the selection matrix (i.e., diagonal matrix with 1 on the selected region and 0 elsewhere). The discrete prolate spheroidal sequences are then the eigenvectors of the concentration matrix $\ma C$ multiplied by $\ma F_W$. This discrete sequence will converge to the continuous-domain solution when the sampling step decreases and the overall support increases. 

If the original selection matrix $\ma S^+$ is substituted by a generalized selection matrix $\ma S = \ma S^+ - \ma S^-$ where $\ma S^+$ and $\ma S^-$ are the selection matrices of the selected and penalized regions respectively, the optimization criterion becomes the generalized optimization criterion that is the topic of this Paper:
\begin{eqnarray}
  \label{eq:gen-slepian-1d-Rayleigh}
   \lambda = \frac{ \vc v^H \ma C \vc v}{\vc v^H \vc v} \text{ with }\ma C = \ma F_W^H\ma S\ma F_W.
\end{eqnarray}
Since all Fourier modes, except the constant mode, come in pairs with the same eigenvalue/frequency, taking an even bandwidth $W$ would mean that one of the pairs is split and therefore the truncated DFT matrix would be ambiguous. Therefore, in this Paper all simulations are done using odd values for the bandwidth.
The matrix $F_W$ can be formed by taking the eigenvectors of the Laplacian of a ring graph with $N$ nodes.

In order to approximate the continuous case with a discrete time simulation, two steps are needed. First, the continuous signals are sampled at sampling frequency $f_s$. If the sampled signal is interpreted as a discrete signal, but still with infinite length, the corresponding frequency domain is limited to the interval $[-\frac{f_s}{2},\frac{f_s}{2}]$, though still continuous. Since it is not possible to run simulations on a signal of infinite length, a finite support of $N$ samples is taken, corresponding to a time duration of $\frac{N}{f_s}$. The Discrete Fourier Transform of the resulting discrete time signal of finite length corresponds to the frequencies 
\begin{equation}
-\frac{f_s}{2} + k\frac{f_s}{N} \text{ with } k=0..N-1.
\end{equation}
Here $f_s$ is always chosen to be 100Hz and $N = 4096$. If the one-sided continuous time bandwidth is chosen to be $F = \alpha 2\pi\frac{f_s}{2}$ with $0<\alpha<1$, the corresponding indices $k$ to be kept are given by:
\begin{equation}
\left\{ k \in \mathbb{N}: \left|-\frac{f_s}{2} + k\frac{f_s}{N}\right| \le \alpha\frac{f_s}{2}\right\}.
\end{equation}
This reverts to taking into account only the first $1+2\left\lfloor\alpha \frac{N}{2}\right\rfloor$ columns of the DFT matrix.
Taking for example $\alpha = 5\%$ yields the following values for $k$:
\begin{equation}
\left\{ k \in \mathbb{N}: \left|-50 + k\frac{100}{4096}\right| \le 2.5\right\} = \{1946..2150\}.
\end{equation}
This means that the first $2150-1946+1 = 205 = 1+2\left\lfloor0.05 \frac{4096}{2}\right\rfloor$ columns of the DFT matrix will be kept in the calculations.

A summary of all parameter values used for the Figures in this work is given in Table~\ref{tab:simulation_params}.

\begin{table}
\centering
\caption{List of parameters for the numerical simulation used in different Figures of this paper, $\Delta_P$ indicates a variable shift for Fig.~6-7.
\label{tab:simulation_params}}
\begin{tabular}{c|c|c|c|c|c|c|c}
\textbf{Fig.}&\textbf{$f_s [Hz]$}&\textbf{$N$}&\textbf{$P_n^+ [s]$}&\textbf{$T_n^+ [s]$}&\textbf{$P_n^- [s]$}&\textbf{$T_n^- [s]$}&\textbf{$\alpha [\%]$}\\\hline
\ref{fig:spectrum_def}&100&4096&2.0&1.0&3.5&0.5&5\\
\ref{fig:slepian-1d}&100&4096&0.5&0.5&2.0&1.0&2\\
\ref{fig:approx_orthogonality}&100&4096&0.5&0.5&2.0&1.0&5\\
\ref{fig:sketch_deltas}&100&4096&2.0&1.0&3.5&0.5&4\\
\ref{fig:interaction}&100&4096&1.5&0.5&$2.5+\Delta_P$&0.5&1\\
\ref{fig:interaction_decay}&100&4096&1.5&0.5&$2.5+\Delta_P$&0.5&1,2,3\\
\ref{fig:reconstruction_example}&100&4096&2.5&1.5&5.5&1.5&1.5\vspace{.5cm}
\end{tabular}
\end{table}


\ifCLASSOPTIONcaptionsoff
  \newpage
\fi



%
\bibliographystyle{IEEEbib}
\bibliography{bibliography}

%

\begin{IEEEbiography}[{\includegraphics[width=1in,height=1.25in,clip,keepaspectratio]{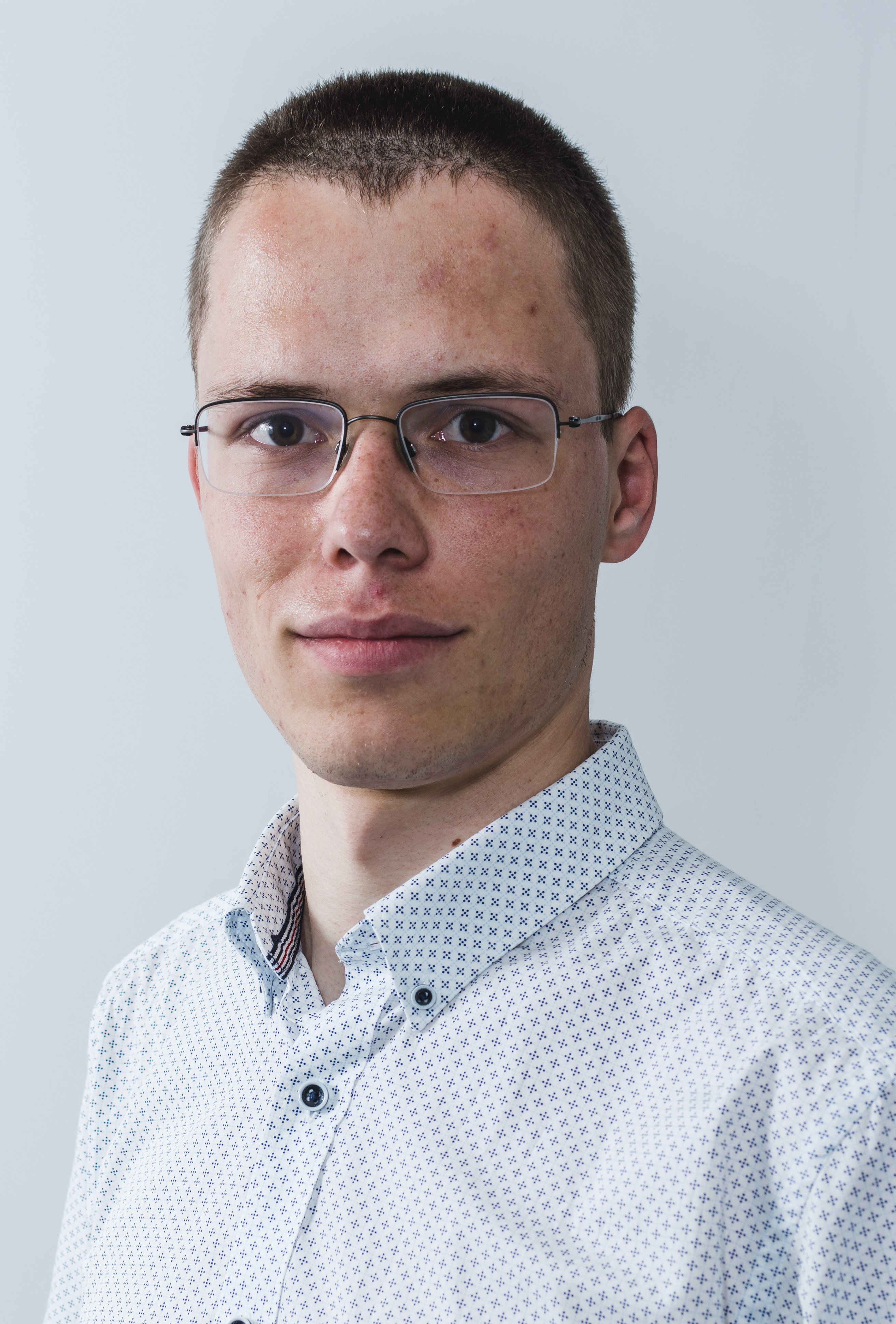}}]{Robin Demesmaeker}
obtained his B.Sc. in Electrical Engineering from Ghent University, Belgium, in 2015 and his M.Sc. in Electrical and Electronic Engineering from the Ecole Polytechnique F\'ed\'erale de Lausanne (EPFL), Switzerland. During his M.Sc. he joined the Medical Image Processing Laboratory (MIP:lab) as a semester project student under supervision of \mbox{Prof. Dimitri Van De Ville} and \mbox{Dr. Maria Giulia Preti}. His main field of interest is the application of information technologies and signal processing in biomedical applications. 
\end{IEEEbiography}


\begin{IEEEbiography}[{\includegraphics[width=1in,height=1.25in,clip,keepaspectratio]{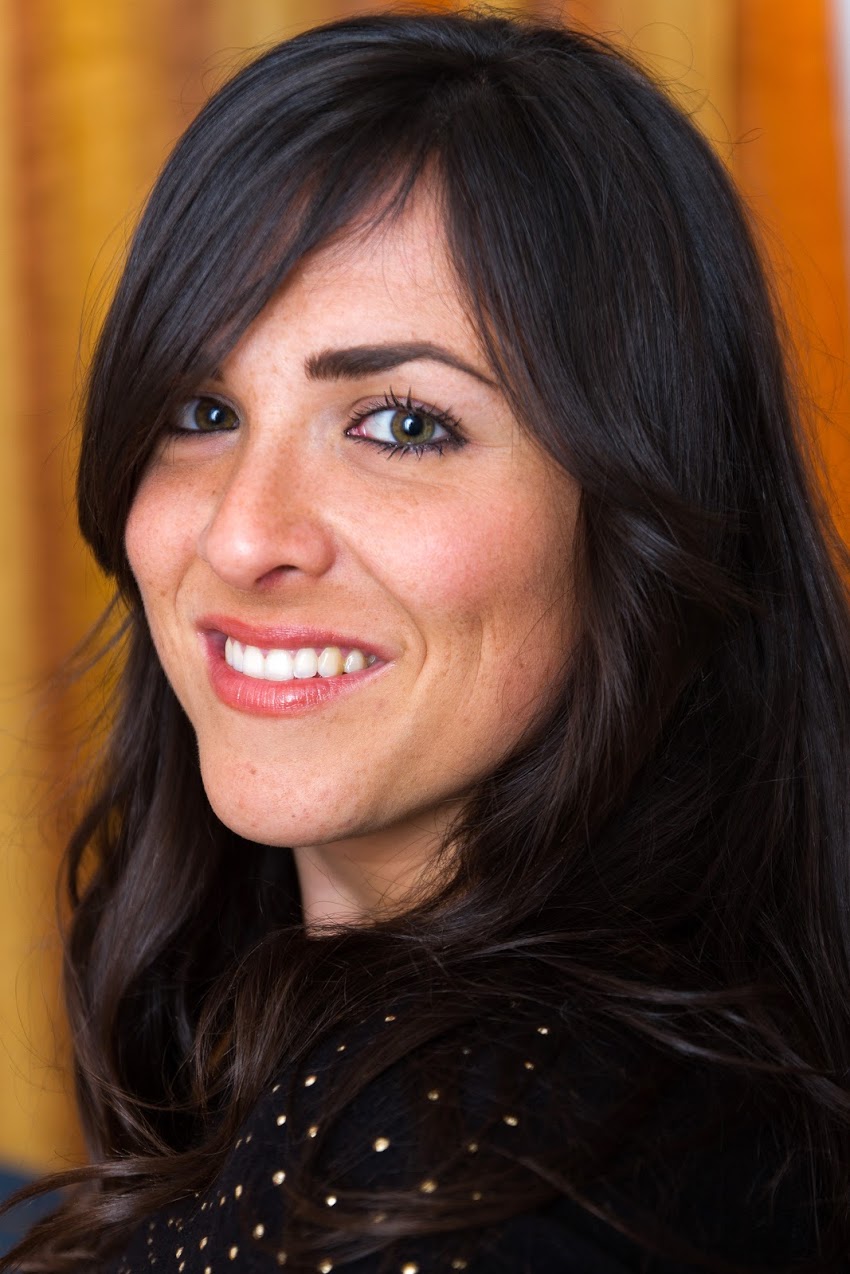}}]{Maria Giulia Preti}
has joined Prof. Van De Ville group as a post-doc in 2013. Her current research aims at understanding the connections between brain functionality and brain microscopic anatomy by using advanced techniques of Magnetic Resonance Imaging.
She received her Ph.D. in Bioengineering at Politecnico di Milano (Milan, Italy) in 2013, after her M.Sc. (2009) and B.Sc. (2007) in Biomedical Engineering, as well at Politecnico di Milano. During her Ph.D., mentored by Prof. Giuseppe Baselli, she focused on advanced techniques of brain magnetic resonance imaging, in particular she developed a method of groupwise fMRI-guided tractography, that revealed to be useful in the in-vivo investigation of the pathophysiological changes across the evolution of Alzheimer's disease. For this project, she had been collaborating full-time with the hospital Fondazione Don Gnocchi in Milan (Magnetic Resonance Laboratory). In 2011, she was awarded a Progetto Rocca fellowship from MIT-Italy and spent a visiting research period at the MIT and Harvard Medical School (Boston, USA), under the supervision of Prof. Nikos Makris.
\end{IEEEbiography}

\begin{IEEEbiography}[{\includegraphics[width=1in,height=1.25in,clip,keepaspectratio]{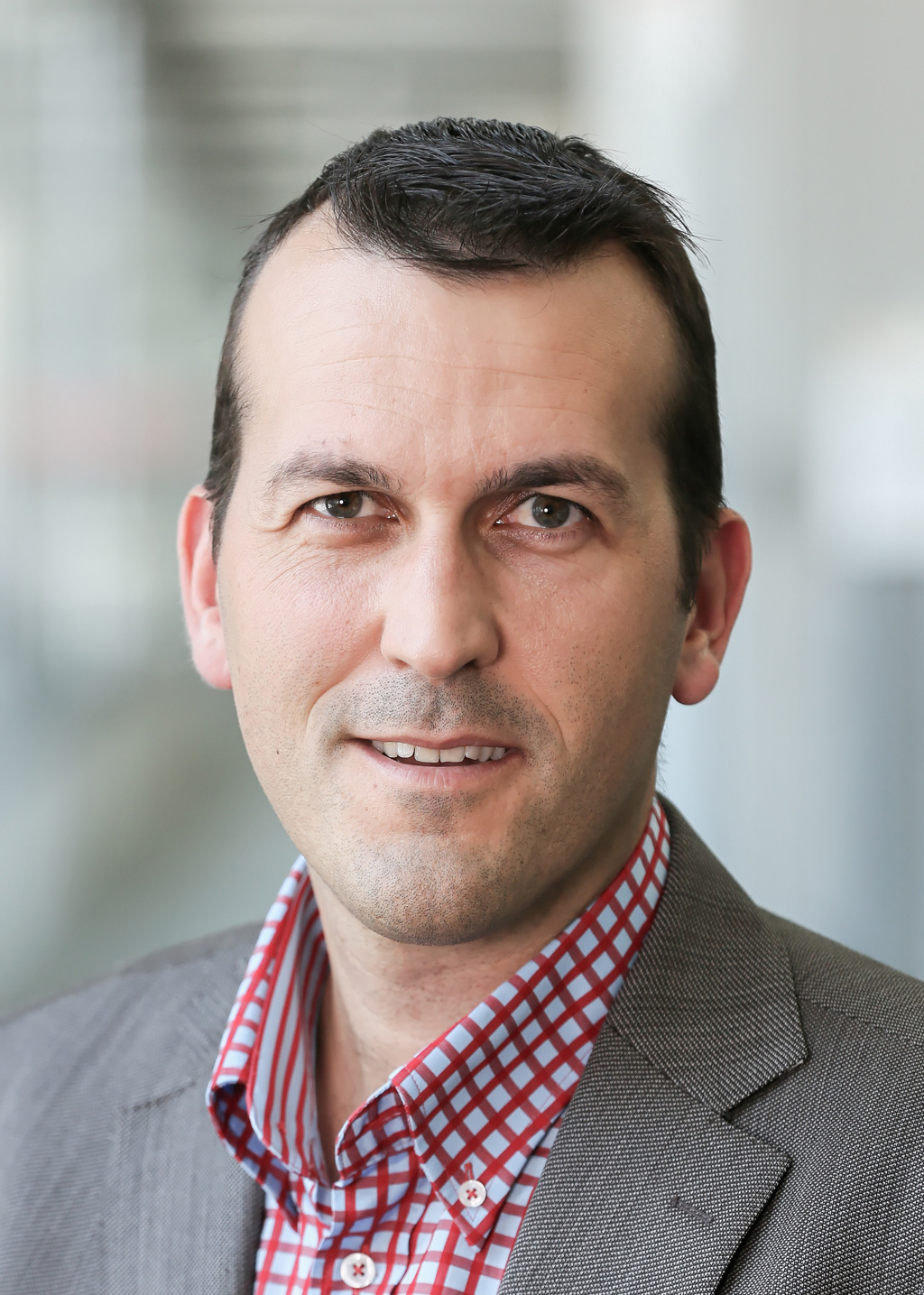}}]{Dimitri Van De Ville} (M'02,SM'12) received the M.S. degree in engineering and computer sciences and the Ph.D. degree from Ghent University, Belgium, in 1998, and 2002, respectively. After a post-doctoral stay (2002-2005) at the lab of Prof. Michael Unser at the Ecole Polytechnique F\'ed\'erale de Lausanne (EPFL), Switzerland, he became responsible for the Signal Processing Unit at the University Hospital of Geneva, Switzerland, as part of the Centre d'Imagerie Biom\'edicale (CIBM). In 2009, he received a Swiss National Science Foundation professorship and since 2015 became Professor of Bioengineering at the EPFL and the University of Geneva, Switzerland. His research interests include wavelets, sparsity, pattern recognition, and their applications in computational neuroimaging. He was a recipient of the Pfizer Research Award 2012, the NARSAD Independent Investigator Award 2014, and the Leenaards Foundation Award 2016.

Dr. Van De Ville served as an Associate Editor for the IEEE TRANSACTIONS ON IMAGE PROCESSING from 2006 to 2009 and the IEEE SIGNAL PROCESSING LETTERS from 2004 to 2006, as well as Guest Editor for several special issues. He was the Chair of the Bio Imaging and Signal Processing (BISP) TC of the IEEE Signal Processing Society (2012-2013) and is the Founding Chair of the EURASIP Biomedical Image \& Signal Analytics SAT. He is Co-Chair of the biennial Wavelets \& Sparsity series conferences, together with V. Goyal and M. Papadakis.
\end{IEEEbiography}




\end{document}